\documentstyle[11pt,epsfig]{article}
\newcommand{\be}{\begin{equation}}
\newcommand{\ee}{\end{equation}}
\newcommand{\ba}{\begin{eqnarray}}
\newcommand{\ea}{\end{eqnarray}}

\newcommand{\ave}[1]{\langle {#1} \rangle}

\begin{document}
\begin{titlepage}
\vspace*{3.5cm}
\begin{center}
\begin{Large}
{\bf In-Medium $\pi-\pi$ Correlations in the '$\sigma$-Meson'-Channel}\\
\end{Large}
\vspace*{0.3cm}
P. Schuck$^{(1)}$,\\
Z. Aouissat$^{(2)}$, F. Bonutti$^{(3)}$, G. Chanfray$^{(4)}$, E.
Fragiacomo$^{(3)}$,  
N. Grion$^{(3)}$ and J. Wambach$^{(2)}$\\
$^{(1)}$ {\small{\it  ISN, 53 avenue des Martyrs,
F-38026 Grenoble C\'edex, France.}}\\
$^{(2)}$ {\small{\it Institut f\"{u}r Kernphysik, Technische Universit\"at 
Darmstadt, Schlo{\ss}gartenstra{\ss}e 9, 64289 Darmstadt, Germany.}}\\
$^{(3)}$ {\small{\it Istituto Nazionale
 di Fisica Nucleare and Dip. di Fisica, 34127 Trieste, Italy.}}\\
$^{(4)}$ {\small{\it IPN-Lyon, 43 Bd. du 11 Novembre 1918,
F-69622 Villeurbanne C\'{e}dex, France.}}\\

\end{center}
\vspace*{2.cm}
\begin{abstract}

\end{abstract}
The present status of the in-medium invariant $\pi-\pi$ mass distribution in the
sigma-meson channel around the threshold is reviewed from the theory
side and contrasted with recent experiments on $(\pi,2\pi)$ knock-out
reactions. A preliminary investigation indicates that the strongly
target-mass dependent invariant mass enhancement of the two
pions can be explained theoretically. A more refined reaction theory is
needed to confirm this result. In the theoretical description, based on
the linear sigma model, emphasis is put on constraints from chiral
symmetry. 
\end{titlepage}
\newpage

In-medium pion-pion correlations have recently attracted much attention both 
on the theoretical and experimental sides. For instance, in the $ J=I=0$ 
(sigma-meson)-channel the interest is quite obvious, since it is common belief 
that some kind of effective sigma meson 
is responsible for the midrange attraction of the nucleon-nucleon potential
\cite{BONN}. It is therefore important to know how such 
a meson is modified in a nuclear medium, a question which is also 
intensively studied for other mesons. In the past we have developed 
phenomenological models in which a 'bare' sigma meson is coupled to the
decay channel into two pions. The coupling constants were adjusted such
that the experimental $\pi\pi$ phase shifts in the $J=I=0$ are reproduced. 
The same procedure was adopted simultaneously for the rho meson in the $I=J=1$ 
channel of the two pions with a perfect reproduction of the position and
the width of the rho meson \cite{CASN}. Certainly, because of the s-wave 
nature of the interaction, the coupling of the sigma meson to the pions is 
much stronger than in the rho-meson channel such that the sigma meson 
becomes completely hybridised with two pions and the mass 
distribution only shows a very broad peak, roughly $500 MeV$ wide.
We then accounted for medium effects by coupling the pions to $\Delta$-h 
and p-h excitations in the usual way and found as a function of density
attractive downward shifts of portions of
the sigma-meson mass distribution, even far below the $2m_{\pi}$ 
threshold \cite{ACS}. This is driving the system into an instability
at densities close to saturation. In-medium vertex corrections which
could give some repulsion in the correlated pion-pair system 
were investigated in ref.~\cite{ZA} and found not to be sufficient 
to hinder this phenomenon.
In fact, it was recognised later that it is absolutely
necessary to fulfil constraints from chiral symmetry in order to 
generate the needed repulsion below threshold which can prevent the 
invasion of the sigma-meson mass distribution into this region.  
Chiral symmetry was implemented phenomenologically in \cite{ARCSW} and it 
was indeed found that the build up of strength below threshold was strongly 
reduced. However, in ref.~\cite{ARCSW} we did not project on the sigma-meson 
channel but rather showed the imaginary part of the $\pi-\pi$ T-matrix  
Because of form factors in the T-matrix, the strength distribution
below threshold is very much suppressed. In the following, we will show  
that the sigma-meson mass distribution proper is still substantially shifted 
downwards in the nuclear medium. However, this will only serve for 
demonstration purposes. For real processes such as {\it eg.} the 
nucleon-nucleon force, mediated by a sigma meson-exchange,
one has to be careful to also include the exchange of a
correlated $\pi-\pi$ pair such that chiral symmetry is conserved \cite{rapp}.\\ 

~As a model for the vacuum $\pi\pi$ s-wave correlations we consider 
first the leading-order contributions of the $1/N$-expansion 
using the two flavour linear $\sigma$-model \cite{ASW}. This, if 
treated correctly, leads to a symmetry-conserving approach and hence 
fulfils  Ward identities and all chiral symmetry constraints \cite{ASW}. 
It also preserves the unitarity of the S-matrix, since
it is based on an RPA-type equation. Therefore, the $\pi\pi$
T-matrix is non-perturbative which is absolutely 
needed if one wants to describe such features as accumulation of
strength and eventually resonances at low energies.\\
Explicitly we have for the T-matrix to leading order
\begin{equation}
T(E, {\vec p}) \,= \frac{V_{\pi\pi}(E, {\vec p})}
{1\,\,-\,\, \frac{1}{2}V_{\pi\pi}(E, {\vec p})\Sigma_{\pi\pi}(E, {\vec p})}~,
\label{eq1}
\end{equation}
where  ${\Sigma}_{\pi\pi}(p)$ is the usual $\pi\pi$ bubble
\\
\begin{equation}
{\Sigma}_{\pi\pi}(p^2) = -i \int\frac{d^4q}{(2\pi)^4}  D_{\pi}(q)
D_{\pi}(p-q)~,
 \label{eq2}
\end{equation}
\\ 
and $V_{\pi\pi}$ the tree-level $\pi\pi$-scattering amplitude given by  
\begin{equation}
V_{\pi\pi}(E, {\vec p})\,=\, N \frac{{\cal E}_{\sigma}^2({\vec 0}) - 
{\cal E}^2_{\pi}({\vec 0})}{ f_{\pi}^2}
\, \frac{ E^2\,-\,{\cal E}_{\pi}^2({\vec p})}
{E^2 \, -\, {\cal E}_{\sigma}^2({\vec p})} ~,
\label{eq3}
\end{equation}
where ${\cal E}_{\pi}^2({\vec q}) = \sqrt{m_{\pi}^2 + {\vec q}^2}$ and 
${\cal E}_{\sigma}^2({\vec q}) = \sqrt{{\cal E}_{\sigma}^2 + {\vec q}^2}$. Here
$m_{\pi}$ is the pion mass and ${\cal E}_{\sigma}$ is the quasi-sigma mass. 
To this order, the pion decay constant $f_{\pi}$ is related 
to the sigma-condensate $\ave{\sigma}$
via: $f_{\pi}^2 = N \ave{\sigma}^2$, where $N$ stands for the number of 
pion charges. \\
To achieve an acceptable description of the s-wave $\pi\pi$ phase shifts we  
supplement the $V_{\pi\pi}$ quasi-potential by form factors such that
\\
\begin{equation}
V_{\pi\pi}(E, {\vec p}; {\vec q}, {\vec q}')\rightarrow v({\vec q})\, 
  V_{\pi\pi}(E, {\vec p})\, v({\vec q}')
 \quad\quad\quad with \quad\quad
v({\vec q})=g\left(1+\frac{q^2}{q_d^2}\right)^{- \alpha}
\label{eq4}
\end{equation}
\\     
This modification does not destroy in any way the properties of chiral
symmetry as is easily verified.
The parameters $g, q_d$ and $\alpha$ are fixed through a fit to the 
data which yields the following values :
$g=0.9, \quad  q_d=1.GeV,  \quad \alpha=3$. The corresponding phase 
shifts are shown in Fig.1. The power $\alpha=3$ of the Yukawa form factor
is somewhat unusual but it accounts for our neglect of u- and t-channel 
exchange contributions and also for the omission of couplings to the 
$K \overline{K}$  channel.\\   
\begin{figure}[hbt]
\centerline{ 
\epsfig{file=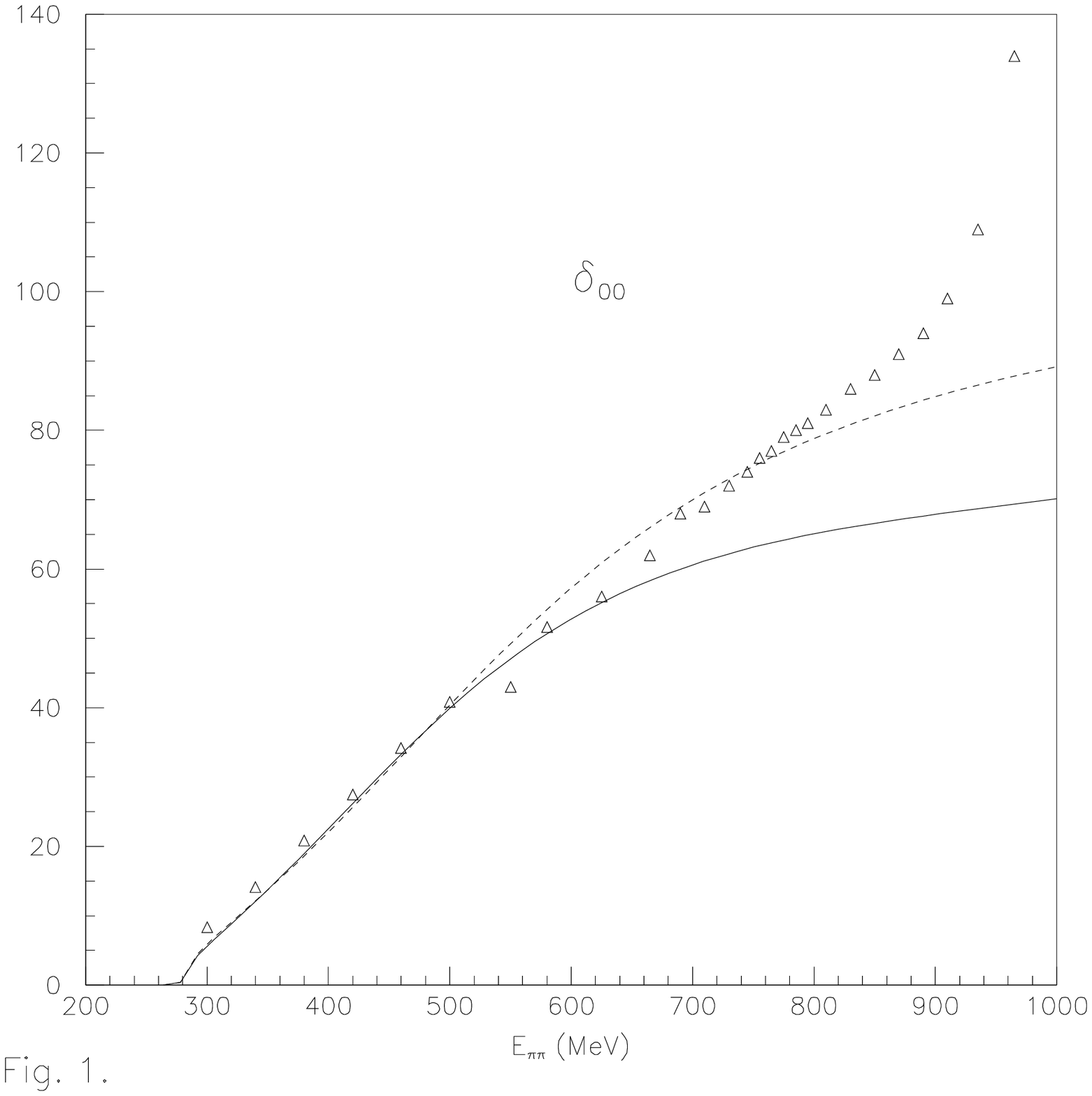,width=12cm,height=9cm,angle=0}}
\caption[fig1]{\small
The s-wave phase shifts for the $\pi\pi$ scattering. Besides the data 
points, the full line denotes
the leading-order result while the dashed line includes t- and 
u-channel corrections.
\label{fig1.} }
\end{figure}
\\
It is interesting to note that the T-matrix in eq.(\ref{eq1}) can also be recast in
the following form
\begin{eqnarray}
T_{ab,cd}(s) \,&=& \delta_{ab}\delta_{cd} 
\frac{D_{\pi}^{-1}(s) - D_{\sigma}^{-1}(s)}{N \ave{\sigma}^2}
\, \frac{D_{\sigma}(s)}{D_{\pi}(s)}~,
\label{eq5}
\end{eqnarray}
\\     
where $s$ is the Mandelstam variable. $D_{\pi}(s)$ and  $D_{\sigma}(s)$ 
are respectively the full pion and  sigma propagators,
while $\ave{\sigma}$ is the sigma condensate. The expression in eq.(\ref{eq5}) 
is in fact a Ward identity which links the $\pi\pi$ four-point function
to the $\pi$ and $\sigma$ two-point functions as well
as to the $\sigma$ one-point function.
To this order, the pion propagator and the sigma-condensate are obtained
from the Hartree-Bogoliubov (HB) approximation \cite{ASW}. 
In terms of the pion-mass $m_{\pi}$ and decay constant 
$f_{\pi}$, they are given by  
\\
\begin{equation}
D_{\pi}(s) = \frac{1}{s - m_{\pi}^2}, \quad\quad \ave{\sigma} =
\frac{1}{\sqrt{N}}f_{\pi}.
\label{eq6}
\end{equation} 
\\
The sigma meson, on the other hand,
is obtained from a Random Phase Approximation (RPA) involving
$\pi-\pi$ scattering \cite{ASW}. It reads    
\\
\begin{equation}
D_{\sigma}(s) \,=\, \left[{ s\,-\, {\cal E}_{\sigma}^2
\,-\, \frac{2 \lambda^4 \ave{\sigma}^2\,{\Sigma}_{\pi\pi}(s)}
{ 1\,-\,  \lambda^2 {\Sigma}_{\pi\pi}(s)}}\right]^{-1}~,
 \label{eq7}
\end{equation}
\\
where $\lambda^2$ is the bare coupling and ${\cal E}_{\sigma}$
is the mean-field sigma mass (mass of the quasi-sigma) given in terms of the
coupling constant, the condensate and the pion mass by: 
\[ {\cal E}_{\sigma}^2 = m_{\pi}^2 + 2 \lambda^2\ave{\sigma}^2.\] 

It is clear from what was said above that the $\sigma$-meson
propagator in this approach is correctly defined since it satisfies a 
hierarchy of Ward identities. 

We now proceed to put the sigma meson in cold nuclear matter. We stress 
again that chiral symmetry is fully preserved. The pion is coupled to 
$\Delta$-h and p-h channels including nuclear short-range correlations simulated 
by the Migdal parameters $g'$ as well as  in-elasticities coming for instance
from the coupling to $2p-2h$ states (see \cite{ACS}). 
The result for various densities is shown in Fig.2.
We see that, as density increases, a strong downward 
shift of the sigma-mass distribution occurs.
However, contrary to earlier phenomenological
models with no repulsion below threshold, the invasion 
of strength below the $E<2m_{\pi}$ threshold region is still strong but  
saturates at around $1.5m_{\pi}$ as density increases. On the other hand, 
we also see that the corresponding imaginary part of the T-matrix is 
less modified.\\
Before discussing the possible relevance of this result for the 
strongly target-mass dependent threshold enhancement of the $\pi\pi$
invariant mass spectrum in recent $\pi, 2\pi$ experiments off nuclei, 
let us briefly discuss some effects which could go in the opposite 
direction. So far, we have only considered self-energy corrections 
to the pions. To be consistent we should consider on the same 
footing vertex corrections which usually go in the opposite 
direction to self-energy effects. Also more care should be given to
the Pauli blocking in matter.
\begin{figure}[hbt]
\centerline{ 
\epsfig{file=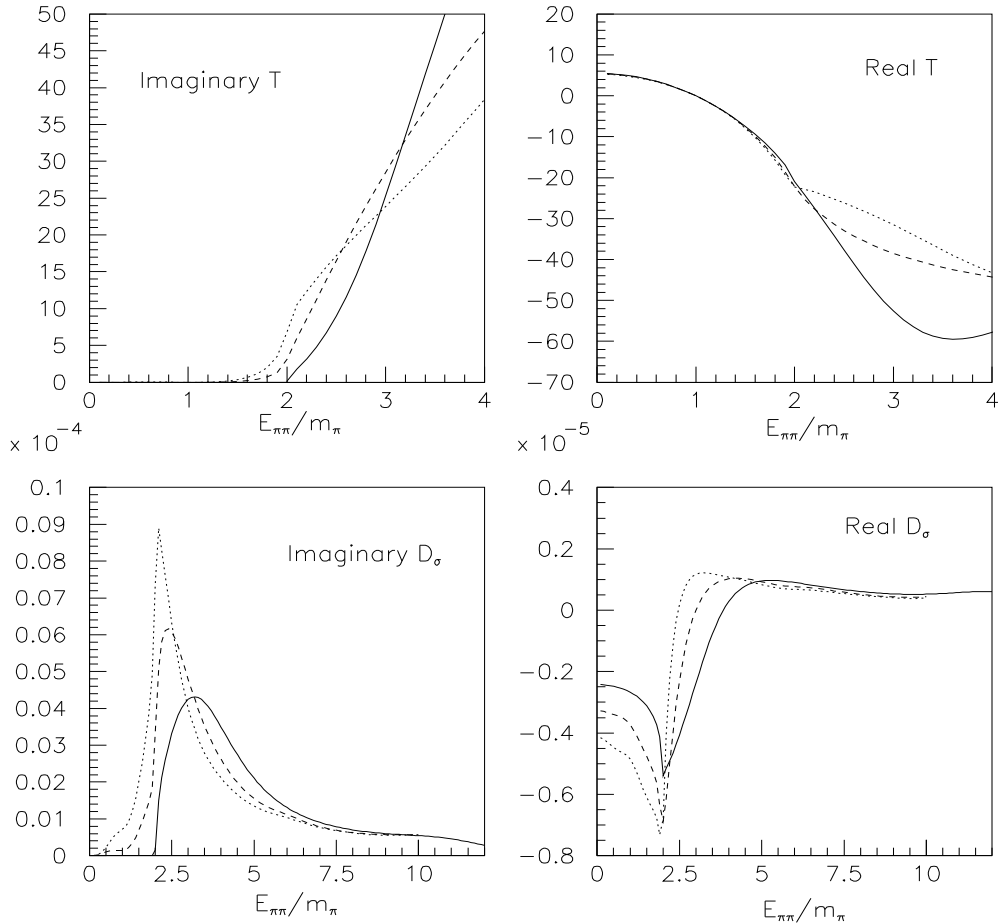,width=15cm,height=14cm,angle=0}}
\caption[fig1]{\small
Results for the leading-order dynamics. The upper left (right) curves denote 
the imaginary (real) part of the $\pi\pi$-T-matrix. 
The lower left (right) curves denote the
imaginary (real) parts of the sigma propagator. 
In all cases, the solid-line, 
the dashed-line and the dotted-line curves are respectively for the vacuum
case, the medium at nuclear density $0.5 \rho_0$ and at normal density.  
\label{fig2.} }
\end{figure}
\\
Indeed, even though the in-medium renormalisation of the single 
pion is well established through the extensive analysis 
of the $\pi$-nucleus optical potential, the 
in-medium renormalisation of correlated pion-pairs 
requires an extra piece  
which is the Pauli exchange contributions to the pion p-h 
and $\Delta$-h self energies.
However, we do not think that these vertex corrections will
completely cancel the effect of accumulation of strength in the 
threshold region as matter density increases. \\

Again we want to mention that in the nuclear medium we have to be 
careful not to consider the sigma isolated from the correlated
pions for instance in their interaction with nucleons. This, in connection 
with chiral symmetry, will be the subject of a separate study.
We only want to demonstrate here that, for the sigma-meson mass 
distribution alone, the in-medium renormalisation is very strong.
On the contrary in the $\pi-\pi$ T-matrix the effect at threshold is 
very much suppressed because of form factors which reflect the
fact that the on-shell T-matrix at threshold has to go to zero 
in the chiral limit(see eq.~(1)). In spite of this there exists, even for the 
T-matrix, a strong reshaping and in particular there is considerable 
strength invading the region below the $2m_{\pi}$ threshold.
It is therefore indeed quite tempting to associate our finding
with the strong strength accumulation found in recent $\pi,2\pi$ knock-out 
experiments off nuclei by Grion et al.\cite{Grion,Bonut}. However  
in these experiments the sigma-meson mass distribution is not 
measured directly but rather the imaginary part of the $\pi\pi$ 
T-matrix.\\

It is clear, from the phase-shift slope, that the s-wave $\pi\pi$ 
scattering
is attractive. The Weinberg scattering length are known to be 
$a_0^0 = \frac{7}{32 \pi} \frac{m_{\pi}}{f_{\pi}^2}$. However,  the 
leading-order contributions of the $1/N$-expansion, as described above, 
leads to too much attraction. The scattering lengths, in the case of
the three physical charges ($N=3$), are given by 
 \[a_0^0 = \frac{9}{32 \pi} \frac{m_{\pi}}{f_{\pi}^2}.\]
Therefore, the t- and u-channel contributions are ultimately needed. They are
known to yield repulsion at threshold and enter as next-to-leading-order 
corrections. 
Before this has been worked out consistently we adopt, for the time being, a more %%@
phenomenological  
quasi-potential picture and use a Lippmann-Schwinger equation as a
unitarization procedure of the full tree-level $\pi\pi$ scattering amplitude
\\
\begin{eqnarray}
V_{ab,cd}(s,t,u) \,&=& \delta_{ab}\delta_{cd} A(s) + 
\delta_{ac}\delta_{bd} A(t) + \delta_{ad}\delta_{bc} A(u)~,\nonumber\\ 
A(s) &=& \frac{m_{\sigma}^2 - m_{\pi}^2}{f_{\pi}^2}
\, \frac{s - m_{\pi}^2}{s - m_{\sigma}^2}~.
\label{eq8}
\end{eqnarray}
\\ 
The functions $A(t)$ and $A(u)$ are obtained from $A(s)$ by substituting
respectively $t$ and $u$ for $s$. This, indeed, corrects for the threshold
and low-energy region and also allows for a reasonable description
of the phase shifts.
For the three-dimensional reduction of the Mandelstam variables $t$ and
$u$ we choose an on-shell one : $t=(\omega_q - \omega_{q'} )^2 -({\vec q} -
{\vec q}')^2$ and $u=(\omega_q - \omega_{q'} )^2 -({\vec q} +{\vec q}')^2$.  
Furthermore the $t$ and $u$ variables will be neglected (in comparison to the
sigma mass) in the denominator
of $A(t)$ and $A(u)$, respectively  This leads to an analytically solvable
integral equation which will account for the full leading-order contribution
of eq.~(\ref{eq1})
and will include a piece of the vertex renormalisation coming from the   
t-and u-channels.
In fact, by this phenomenological procedure, the $A(t)$ and $A(u)$ pieces 
play a role, similar to the Migdal parameters for the pion self-energy
in the nuclear medium, which account for the repulsion
present in the vertex corrections.    
This modification still preserves the symmetry conserving 
properties of the iterated T-matrix, as is easily verified. As a regularisation
of the divergent integrals, we supplement the T-matrix in eq.~(\ref{eq5}) with
a monopole-like form factor with an adjustable cutoff parameters $Q_d$.   
For a bare sigma mass, $m_{\sigma} =1. GeV$, and a cutoff, $Q_d = 8 m_{\pi}$,
one gets a reasonable description of the phase shifts up to $800 MeV$ (Fig.1).
The corresponding imaginary part of the T-matrix is shown in Fig.3. \\
\begin{figure}[hbt]
\centerline{ 
\epsfig{file=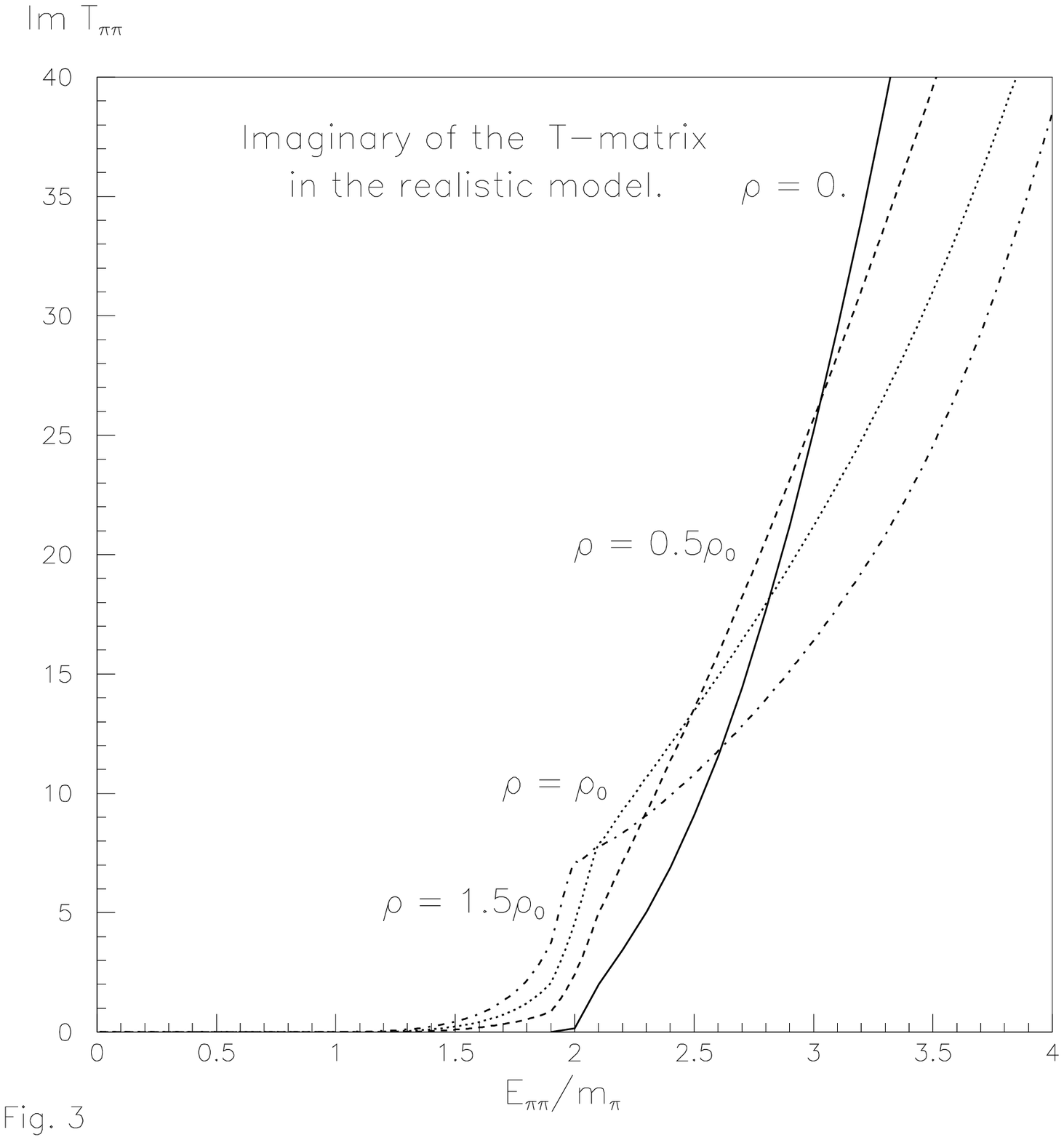,width=14cm,height=9cm,angle=0}}
\caption[fig1]{\small
The imaginary part of the T-matrix with t- and u-channel corrections. 
The in-medium modifications are accounted for by means of a realistic model. 
\label{fig4.} }
\end{figure}

As a first attempt to explain with our theory the findings of 
Bonutti et. al. \cite{Bonut} we consider the scenario of quasi-free
scattering.
That this process dominates the pion knock-out reaction has  been 
demonstrated experimentally in measuring a proton in coincidence
with the pions \cite{pluta}. The elementary process has
actually been calculated by Oset and Vicente-Vacas \cite{oset} 
using a perturbative treatment at the tree level.  
For the invariant
$\pi\pi$ mass , this theory explains reasonably well the $^2H$ data 
 (see \cite{oset}). 
However, the mass dependence of the cross section and notably the resonant 
structure at threshold, growing with target mass, remains totally
unexplained.   
As an attempt to include in-medium effects in this process, 
the authors of ref. \cite{oset} considered 
the in-medium modification of the one-pion-exchange pole-graph
which was found to be negligible. 
It is clear from what was said above that the crucial region where one should 
look for possibly important in-medium effects is the $2m_{\pi}$ threshold. 
On the other hand this region seems to be 
singled out by the in-medium $\pi\pi$ correlation processes.
Therefore it is rather natural to consider those
as a final state interaction effects in describing the 
($\pi$, $2\pi$) knock-out data.       

Therefore, as a first step, we propose to include these final state interactions 
using our model for the in-medium $\pi\pi$ correlations, in the analysis
performed by Oset\footnote[1]{We are grateful to E. Oset and M. J. Vicente
Vacas for providing 
to us with their code.} et al. in ref.\cite{oset}. This will be done in a very 
schematic
way in order to obtain some insight as to whether or not our theory can at all 
explain the features found in these pion knock-out reactions.  
We thus replace in the analysis of ref.\cite{oset} the tree-level
four-pion vertex  by our
medium-modified $\pi\pi$-$T$-matrix which, as we have said, gives 
rise to the $\sigma$-meson mass distribution shown earlier. 
Furthermore, to keep the numerical calculation within reasonable limits, we 
use a rather crude model for the in-medium renormalisation of the pions. Since  
the latter are subject to particle-hole and delta-hole couplings with  
p-wave dominance, one can show by comparison with more realistic studies that,
to a good approximation, the single-pion dispersion relation 
in matter can be modelled in the following way
\[\omega_{\pi} ({\vec q}) = \sqrt{m_{\pi}^2 + {\vec q}^2 }\quad \rightarrow 
\quad \omega_{\pi}^{\gamma}({\vec q}) = \sqrt{m_{\pi}^2 + \gamma {\vec q}^2 }~, \]
where $\gamma$ takes values from $0.8$ to $0.4$ depending on  
density. The results for the in-medium strength distribution 
in this toy-model approach is shown in Fig.~4 for typical values of the parameter
$\gamma$.\\
By comparing the strength distributions from the full (realistic) model of  
Fig.~3 and those from the toy model in Fig.4 one clearly sees that the toy model
is overestimating the effects near threshold. On the other hand, the toy
model misses completely the subthreshold strength. In fact, in the full
model, the presence of subthreshold  p-h and 2p-2h cuts renders the
peaks at threshold much broader than in the toy model. Therefore the strength
which is accumulated at threshold in the case of the toy model is spread into the 
subthreshold region in the case of the full model.\\
\begin{figure}[hbt]
\centerline{ 
\epsfig{file=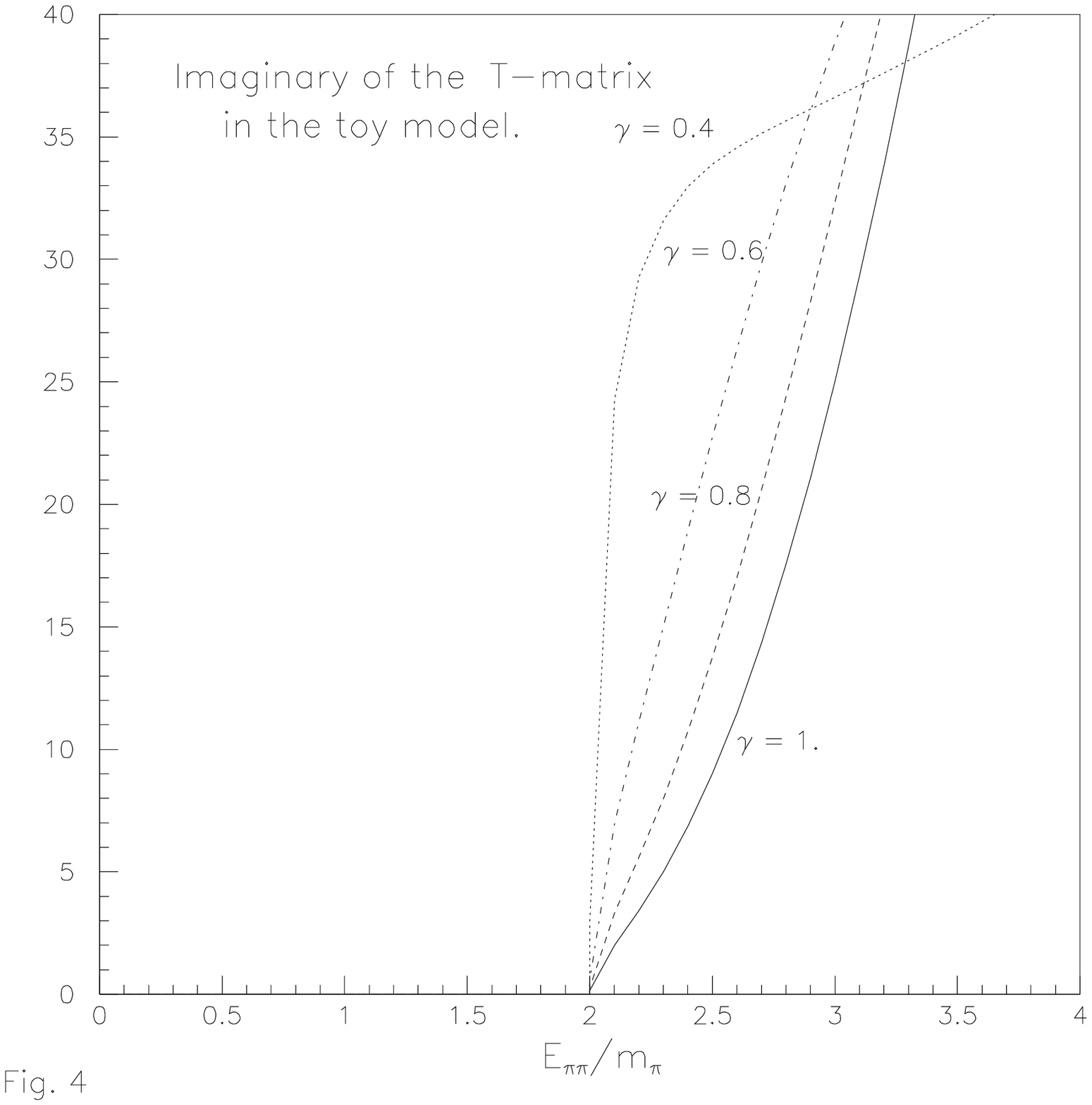,width=14cm,height=9cm,angle=0}}
\caption[fig1]{\small
The imaginary part of the T-matrix with t- and u-channel corrections. 
In-medium modifications are accounted for by means of a toy model. 
\label{fig5.} }
\end{figure}
At first glance, the use of the toy model  to assess the  
in-medium effects on the ($\pi$, $2\pi$) knock-out reaction off nuclei,
may seem rather erroneous due to this overestimation of the strength 
at threshold.
However, one should keep in mind that a reaction theory for the knock-out 
process has to take into account the finite three-momenta of the emitted 
pion-pairs, whereas the calculation presented in Figs.~ and 4 are all 
done in a
back-to-back kinematics (c.m frame).  Finite three-momenta ${\vec P}$ of 
the pair induce
in fact an interesting feature which we want to discuss now.\\ 
Consider two pions detected  outside the nucleus at 
three-momenta  ${\vec q}_1$, ${\vec q}_2$ 
(with ${\vec P}= {\vec q}_1 + {\vec q}_2$)
and energies~    
$\omega_{\pi} ({\vec q}_1) = \sqrt{m_{\pi}^2 + {\vec q}_1^2 }$,~ 
$\omega_{\pi} ({\vec q}_2) = \sqrt{m_{\pi}^2 + {\vec q}_2^2 }$,   
  respectively.
Inside the nucleus,
on the other hand, this pair of pions have three-momenta 
${\vec k}_1$, ${\vec k}_2$ and
 energies $\omega_{\pi}^{\gamma}({\vec k}_1)$, 
 $\omega_{\pi}^{\gamma}({\vec k}_2)$
 related  by the in-medium dispersion 
\[\omega_{\pi}^{\gamma}({\vec k}_1) =
 \sqrt{m_{\pi}^2 + \gamma {\vec k}_1^2 }~,\quad 
 \omega_{\pi}^{\gamma}({\vec k}_2) =
 \sqrt{m_{\pi}^2 + \gamma {\vec k}_2^2 }~.
 \]   
Knowing that the energy of each particle, in either the vacuum or the
medium, has to be the same: 
$\omega_{\pi} ({\vec q}_1) = \omega_{\pi}^{\gamma}({\vec k}_1)$ and 
$\omega_{\pi} ({\vec q}_2) = \omega_{\pi}^{\gamma}({\vec k}_2)$, one can
one can extract a scaling relation between the three-momenta of each particle
inside and outside the nucleus  
\[ {\vec q}_1 = \sqrt{\gamma}\,  {\vec k}_1 ~,\quad 
 {\vec q}_2 = \sqrt{\gamma}\, {\vec k}_2 ~. \]   
Hence, the invariant mass $\tilde{M}_{\pi\pi}$ of a pion-pair inside 
the nucleus  which is given
by
\[ \tilde{M}_{\pi\pi} = 
( \omega_{\pi}^{\gamma}({\vec k}_1) + \omega_{\pi}^{\gamma}({\vec k}_1))^2\,
-\, ( {\vec k}_1 + {\vec k}_2 )^2 ~,\]  
is related to the invariant mass, $M_{\pi\pi} = 
( \omega_{\pi}({\vec q}_1) + \omega_{\pi}({\vec q}_1))^2\,
-\, ( {\vec q}_1 + {\vec q}_2 )^2$,  
of the same pion-pair outside the nucleus
through the relation :
 \[ \tilde{M}_{\pi\pi} =  M_{\pi\pi} - \left( \frac{1}{\gamma} - 1\right) \vec{
 P}^2~. \] 
One can see that, for the typical values taken by the $\gamma$ parameter 
($\gamma \le 1$), 
 the pion-pair, observed at the invariant mass  $M_{\pi\pi}$,
has actually a lower in-medium invariant mass 
$\tilde{M}_{\pi\pi} \le M_{\pi\pi}$. This also suggests that, for finite
three-momenta pion-pairs (${\vec P} \neq {\vec 0}$), one is in fact able to  
probe the strength of the $\pi\pi$ mass
distribution at an in-medium invariant mass even below the $2 m_{\pi}$ 
threshold. 
This interesting feature is not possible for the case of 
back-to-back pions even though each individual pion has a modified in medium
dispersion relation owing to the factor $\gamma$.  
Therefore, since we presently do not possess a consistent reaction-theory
at finite total three-momentum, it may be advantageous not to use the full realistic 
model of Fig.~3  in which the strength is totally spread out into the region
below threshold but
instead  use the toy model which keeps the total strength concentrated
at threshold. This option may eventually compensate 
for the inadequate back-to-back kinematics inherent in our model for in-medium
$\pi\pi$ correlations. Of course, this argument certainly needs  better 
support from a quantitative analysis at finite total three-momentum.  This
study is presently underway.

Before concluding this short note, we present a numerical calculation going
along the lines sketched above. To the theoretical analysis by Oset and 
Vicente Vacas of the $(\pi, 2\pi)$ knock-out reaction off nuclei, we add 
$\pi\pi$ final state interactions. This is done by replacing the tree-level
$\pi\pi$-vertex considered in \cite{oset} by an in-medium renormalised
$\pi\pi$  T-matrix. As indicated above this is only done in the
framework of the economical toy model. 

\begin{figure}[hbt]
\centerline{ 
\epsfig{file=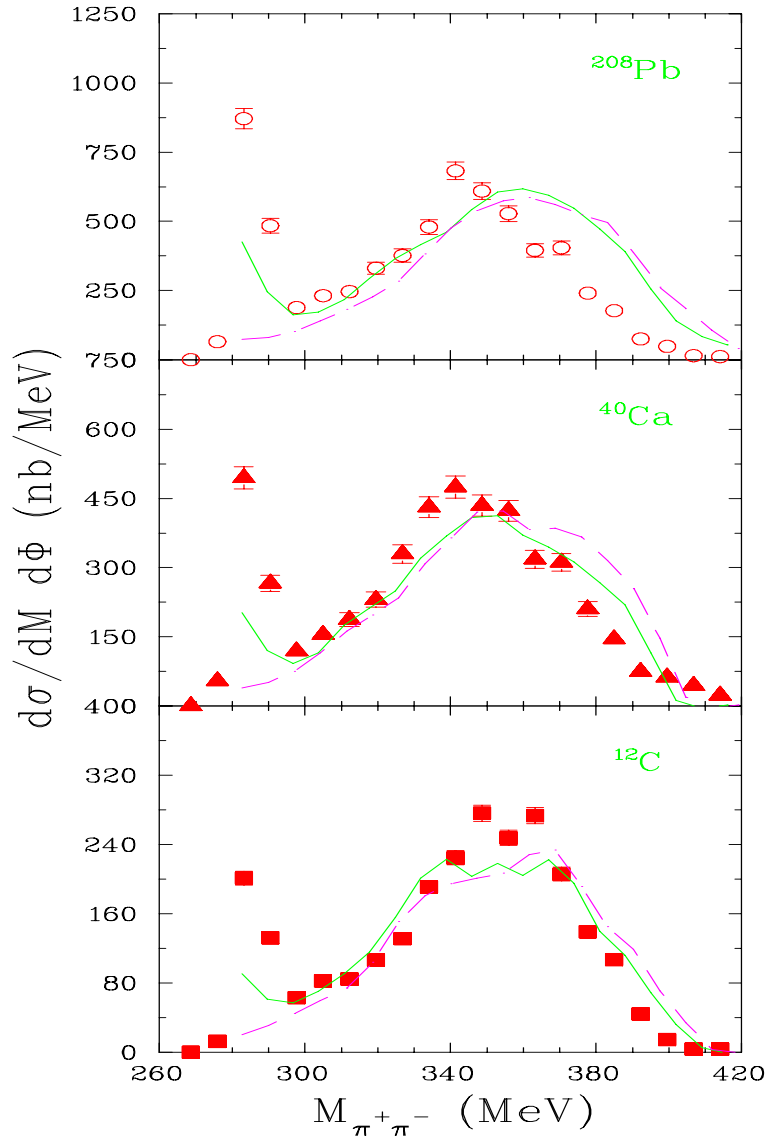,width=10cm,height=15cm,angle=0}}
\caption[fig1]{\small
The $\pi^+\pi^-$ invariant-mass distribution 
in the $(\pi, 2\pi)$ knock-out reaction off Carbon, Calcium and Lead.
Along with the experimental points curves from the theoretical
calculations  of Oset and Vicente Vacas \cite{oset} are displayed 
as dashed-dotted lines. The full lines 
include the in-medium final-state interaction in the $\pi\pi$ vertex.
The in-medium modifications are taken into account by means of the toy model
as discussed in the text.         
\label{fig5.} }
\end{figure}

We can see from the theoretical curves that the experimental findings can 
roughly be described with values of the $\gamma$ parameter in the range of
0.6 to 0.8 (see figure 5). 
As discussed above, this essentially stems from the fact that the in-medium 
renormalisation of the pions induces an important
downward shift of the strength in the $\pi\pi$ T-matrix. This analysis
was made to illustrate the basic idea. It will be important to 
look carefully into vertex corrections which will certainly
have a moderating effect. This remains to be seen in future work.

At the end let us emphasise that the low invariant $\pi-\pi$ mass accumulation 
considered here is of different origin than a similar effect obtained from $A(p, 2\pi)X$
reactions at SATURNE \cite{pluta,saturn}. There the
incident proton energy was $1.6 GeV$ such that the pions have much higher
energies than in the TRIUMF experiment. Because of that the pions in the SATURNE
experiment have a much shorter mean free path, i.e. they are strongly absorbed
by the medium, leading to a pronounced shadowing effect. The collinearity of the 
pions leaving the nucleus on the nuclear-matter-free side induces a low
invariant-mass enhancement of the $\pi\pi$ cross section which increases with the
size of the target nucleus. This scenario was confirmed in \cite{pluta} from 
numerical BUU simulation but also experimentally from the fact that the cross
section for heavier targets is maximum for small opening angles of the pions.
In contrast, at TRIUMF, the outgoing pions are of low energy and therefore have
a long mean free path which is confirmed by the fact that the cross section is
flat as a function of the opening angles (see reference \cite{Grion}).
This kind of scenario is also individually confirmed by the mass dependence 
of the total $(\gamma, 2\pi)$ production cross section on nuclei which, as a
function of mass, is totally flat for low energy $\gamma$'s $( 400 MeV)$
whereas it decreases for high energy photons (1 GeV).  We therefore
think that the experiment analysed here has nothing to do with nuclear
shadowing and that the observed mass enhancement really signals a collective 
effect resulting in a strong downward shift of part of the $\pi^+-\pi^-$-and
$\sigma$-meson strength distribution.\\

\noindent\underline {\bf Acknowledgements} :

One of us (Z.A)  acknowledges financial support from 
GSI-Darmstadt and would like to thank all the group members of INFN-Trieste
for their warm hospitality.\\

\end{document}